\documentclass[twocolumn,showpacs,pra,aps,superscriptaddress]{revtex4}
\usepackage{graphicx}
\usepackage{dcolumn}
\usepackage{bm}
\usepackage{hyperref} 

\begin{document}

\title{Transient effects and reconstruction of the energy spectra in \\
the time evolution of transmitted Gaussian wave packets}

\author{Sergio Cordero}
\author{Gast\'on Garc\'{i}a-Calder\'{o}n}
\altaffiliation[Corresponding author.\,\,\,]{gaston@fisica.unam.mx}
\affiliation{Instituto de F\'{i}sica, Universidad Nacional Aut\'{o}noma de M\'{e}xico, Apartado postal 20-364,
M\'{e}xico 01000, Distrito Federal, Mexico}

\date{\today}

\begin{abstract} We derive an exact analytical solution to the  time-dependent Schr\"odinger equation for transmission of a Gaussian wave packet through  an arbitrary potential of finite range. We consider the situation where the initial Gaussian wave packet is sufficiently broad in momentum space
to guarantee that the resonance structure  of the system is included in the dynamical description. 
We demonstrate that the transmitted wave packet exhibits a transient behavior which  at very 
large distances and long times may be written as the free evolving Gaussian wave packet solution
times the transmission amplitude of the system and hence it reproduces the resonance spectra of the system. 
This is a novel result that predicts the ultimate fate of the transmitted Gaussian wave packet.
We also prove that at a fixed distance and very long times the solution goes as $t^{-3/2}$ which
extends to arbitrary finite range potentials previous analysis on this issue. Our results are exemplified for 
single and multibarrier systems.
\end{abstract}


\maketitle

\section{Introduction}

The transmission of wave packets through one-dimensional potentials is a model that has been of great relevance both from a pedagogical point of view, as discussed in many quantum mechanics textbooks, and in research, particularly since the advent of artificial semiconductor quantum structures \cite{ferry,mizuta}. There are studies on the dynamics of tunneling \cite{gcr97,pereyra03,konsek,andreata,fu,granot,wulf} or  work on controversial issues, as the tunneling time  problem \cite{maccoll,landauer,muga08,gcv03}, and on related topics as the Hartman effect \cite{hartman,muga03}
or the delay time \cite{bohm,muga02,hgc03}. Most time-dependent numerical studies consider  Gaussian
wave packets as initial states \cite{konsek,mizuta,hauge91,harada}, though  in some recent work, the formation of a quasistationary state in the scattering of wave packets on finite one-dimensional periodic structures involves also some analytical considerations \cite{rusos}.
Analytical approaches have been mainly concerned with cutoff quasi-monochromatic initial states in a quantum shutter setup \cite{moshinsky52,holland,gcr97,gcvy03,hgc03}. In recent work, however, analytical solutions to the time-dependent wave function have been discussed using initial  Gaussian wave packets for square barriers 
\cite{muga03,vrc07}, delta potentials \cite{andreata,vrc07} and resonant tunneling systems near a single 
resonance \cite{wulf}.

We obtain an exact analytical solution to the  time-dependent Schr\"odinger equation for transmission of an initial Gaussian wave packet through an arbitrary potential of finite range. We refer to  the physically relevant case where the initial Gaussian wave packet is sufficiently far from the interaction region so that the corresponding tail near that region is very small and hence may be neglected. Since the infinite limit of a very broad cutoff Gaussian wave packet in configuration space, \textit{i.e}, that leading to a cutoff plane wave, has been discussed analytically elsewhere \cite{gcr97}, we focus the discussion here to cases where the initial cutoff Gaussian wave packet is sufficiently broad in momentum space so that all the resonances of the quantum system are included in the dynamical description.
We demonstrate that the profile of the transmitted wave packet exhibits a transient behavior
which at very large distances and long times may be expressed as the free evolving wave packet modulated by
the transmission amplitude of the system. To the best of our knowledge this is a novel and interesting result.
We also analyze the transmitted solution at a fixed distance away from the potential  at very long times, and find that
it behaves as $t^{-3/2}$. Our result generalizes to arbitrary potentials of finite range previous analysis
involving specific potentials models and numerical calculations \cite{mdsprb95}.

The paper is organized as follows. In section II, some relevant aspects of the formalism of resonant states are reviewed and a formal solution for the transmitted pulse is given as an expansion in terms of these states and the corresponding complex resonance poles. In Section III the analytical expression for the transmitted Gaussian wave packet is derived and 
some limits are discussed. Section IV refers to some examples, specifically single barrier and double and quadruple barrier resonant tunneling systems are considered and a subsection provides some remarks concerning the tunneling time problem.
Section V gives the concluding remarks and, finally, the Appendixes discuss, respectively, the analysis of the effect of 
the cutoff in the solution, and a general method to calculate the complex poles of the transmission amplitude.

\section{Resonance expansion of the time-dependent solution}

Let us consider the time evolution of an initial state $\psi(x,0)$ of a particle of energy $E_0=\hbar^2 k^2_0/2m$,
approaching from $x<0$  a potential $V(x)$ that extends along the interval $0 < x < L$. The time-dependent solution
along the transmitted region $ x \geq L$  reads \cite{muga96,gcvy03}
\begin{equation}
\psi(x,t)= \frac{1}{\sqrt{2\pi}}\int_{-\infty}^{\infty} dk\, \phi_0(k){\bf t}(k)e^{ikx-i\hbar k^2t/2m}
\label{1}
\end{equation}
where ${\bf t}(k)$ is the transmission amplitude of the problem and $\phi_0(k)$ is the Fourier transform of the initial function $\psi(x,0)$.

One may write the transmission amplitude in terms of the outgoing Green function $G^+(x,x';k)$ of the problem as
\begin{equation}
\mathbf{t}(k) = 2ikG^+(0,L;k)e^{-ikL}.
\label{2}
\end{equation}
It is well known, that the function  $G^+(x,x';k)$, and hence the transmission amplitude ${\bf t}(k)$, possesses an infinite number of complex poles $\kappa_n$, in general simple,  distributed on the complex $k$ plane in a well known manner \cite{newton}. Purely positive and negative imaginary poles $\kappa_n \equiv i\gamma_n$ correspond, respectively, to bound and antibound (virtual) states, whereas complex poles are distributed along the lower half of the $k$ plane. We denote the complex poles on the fourth quadrant by $\kappa_n = \alpha_n-i\beta_n$. It follows from time reversal considerations \cite{rosenfeld} that those on the third quadrant, $\kappa_{-n}$, fulfill $\kappa_{-n}=-\kappa_n^*$. The complex poles may  be calculated by using iterative techniques as the Newton-Raphson method \cite{raphson}, as discussed in the Appendix. Usually one may obtain a resonance expansion for ${\bf t}(k)$ by expanding $G^+(0,L;k)$ in terms of its complex poles and residues \cite{gcrr93}. Here we find more convenient to expand instead $G^+(0,L;k)\exp(-ikL)$ to obtain,
\begin{equation}
{\bf t}(k) = ik\sum_{n=-\infty}^{\infty} \frac{r_n}{k-\kappa_n}e^{-i\kappa_nL},
\label{3}
\end{equation}
where the residues $r_n$ are given by
\begin{equation}
r_n=\frac{u_n(0)u_n(L)}{\kappa_n}.
\label{res}
\end{equation}
The functions $u_n(x)$ appearing in  Eq. (\ref{res}) satisfy the Schr\"odinger equation to the problem with complex eigenvalues $E_n=
\hbar^2\kappa^2_n/2m$ =$\mathcal{E}_n-i\Gamma_n/2$ and obey the purely outgoing boundary conditions
\begin{equation}
\left [\frac{d}{dx} u_n(x)\right ]_{x=0} = -i\kappa_nu_n(0),\, \left [\frac{d}{dx} u_n(x)\right ]_{x=L} = i\kappa_nu_n(L)
\label{bc}
\end{equation}
normalized according to the condition \cite{gcr97},
\begin{equation}
\int_0^L u_n^2(x)dx + i \frac{u^2_n(0)+u^2_n(L)}{2\kappa_n}=1.
\label{norm}
\end{equation}
Substitution of Eq. (\ref{3}) into Eq. (\ref{1}) yields,
\begin{widetext}
\begin{equation}
\psi(x,t)= \frac{i}{\sqrt{2\pi}}\sum_{n=-\infty}^{\infty} r_ne^{-i\kappa_nL}
\int_{-\infty}^{\infty}dk\,\frac{k}{k-\kappa_n}\phi_0(k)e^{ikx-i\hbar k^2t/2m}
\label{1aa}
\end{equation}
\end{widetext}

It is convenient to make use of the identity $k/(k-\kappa_n) \equiv 1 + \kappa_n/(k-\kappa_n)$
to rewrite the solution given by Eq. (\ref{1aa}) as
\begin{equation}
\psi(x,t) = C\psi^f(x,t) + \sum_{n=-\infty}^{\infty} \psi_n(x,t),
\label{3ap}
\end{equation}
where $C$ is a constant that depends only on the potential through the values of the $r_n$'s and $\kappa_n$'s,
\begin{equation}
C=i\sum_{n=-\infty}^{\infty} r_ne^{-i\kappa_nL},
\label{coefc}
\end{equation}
$\psi^f(x,t)$ stands for  the free wave packet solution
\begin{equation}
\psi^f(x,t)= \frac{1}{\sqrt{2\pi}}\int_{-\infty}^{\infty} dk\, \phi_0(k)e^{ikx-i\hbar k^2t/2m},
\label{3bp}
\end{equation}
and  $\psi_n(x,t)$  is given by
\begin{eqnarray}
\psi_n(x,t) &=& ir_nk_ne^{-i\kappa_nL} \times \nonumber \\[.3cm]
&& \int_{-\infty}^\infty \frac{dk}{\sqrt{2\pi}}\frac{\phi_0(k)}{k-\kappa_n}e^{ikx-i\hbar k^2t/2m}.
\label{3cp}
\end{eqnarray}
\section{Analytical solution for a cutoff Gaussian pulse}

Consider now a particle described initially by a cutoff Gaussian wave packet
\begin{equation}\label{5}
\psi(x,0) = \left\{ \begin{array}{l l} A_0e^{-(x-x_c)^2/4\sigma^2}
e^{ik_0x}, & x<0 \\[.5cm] 0, & x>0
\end{array} \right.,
\end{equation}
where $A_0$ is the normalization constant,  $x_c$, $\sigma$ and $k_0$ are, respectively, the center, the effective
width and the wavenumber corresponding to the incident energy $E_0$ of the wave packet.

In order to calculate the free evolving wave packet given by Eq. (\ref{3bp}) and the integral term on the right-hand side of Eq. (\ref{3cp}),  one needs to know the Fourier transform of  the Gaussian cutoff wave packet. This is given
by \cite{vrc07},
\begin{equation}
\phi_0(k) = A_0 \omega(iz),
\label{8}
\end{equation}
where
\begin{equation}
A_0=\frac{1}{\sqrt{2\pi}}\left[\frac{(2\pi)^{1/4}\sigma^{1/2}}{\sqrt{\omega(iz_0)}}\right]
\label{a0}
\end{equation}
where $z$ and $z_0$  are given by
\begin{equation}
z = \frac{x_c}{2\sigma} - i (k-k_0)\sigma,
\label{z}
\end{equation}
and
\begin{equation}
z_0 = \frac{x_c}{\sqrt{2}\,\sigma},
\label{z0}
\end{equation}
and $\omega(iz)$ is the Faddeyeva function \cite{faddeyeva,abramowitz}.

Let us place the initial wave packet along the region  $x_c < 0$. As pointed out above, here we shall be concerned
with the  physically relevant  situation where the tail of the initial Gaussian wave packet is very small near the
interaction region. It is then convenient to consider the symmetry relationship of the Faddeyeva function
\cite{faddeyeva,abramowitz},
\begin{equation}
\omega(iz) =  2e^{z^2}-\omega(-iz)
\label{8aba}
\end{equation}
and follow an argument given be Villavicencio \textit{et.al.} for the free and $\delta$ potential cases \cite{vrc07}.
These authors obtain that provided
\begin{equation}
\left |\frac{x_c}{2\sigma} \right | \gg 1; \qquad x_c < 0,
\label{8bb}
\end{equation}
one may approximate $\omega(iz)$ as
\begin{equation}
\omega(iz) \simeq  2e^{z^2}.
\label{omega}
\end{equation}
In Appendix A we show that the above approximation holds also for the general case of finite range potentials.
Using Eqs. (\ref{8}) and (\ref{omega}) into Eq. (\ref{3bp}) leads to an analytical expression
for the free evolving cutoff Gaussian wave packet \cite{vrc07}, that we denote by $\psi^a_f(x,t)$, that is identical to
the exact analytical expression for an extended initial gaussian wave packet \cite{vrc07},
\begin{eqnarray}
\psi^a_f(x,t)&=&\frac{1}{(2\pi)^{1/4}}\frac{1}{\sigma^{1/2}} \frac{e^{i(k_0x-\hbar k_0^2t/2m)}}{\sqrt{1+it/\tau}}
\times \nonumber \\ [.3truecm]
&&\exp \left \{- \frac{[x-x_c-(\hbar k_0/m) t]^2}{4\sigma^2 \left [ 1 +it/\tau \right ]} \right\},
\label{free}
\end{eqnarray}
where
\begin{equation}
\tau=\frac{2m \sigma^2}{\hbar}.
\label{tau}
\end{equation}

Let us now substitute Eq. (\ref{omega}) into the integral term in Eq. (\ref{3cp}) to obtain
\begin{eqnarray}
\psi^a_n(x,t) &=& ir_n\kappa_ne^{-i\kappa_nL} \times \nonumber \\[.3cm]
&& 2A_0\int_{-\infty}^\infty \frac{dk}{\sqrt{2\pi}}\frac{e^{z^2}}{k-\kappa_n}e^{ikx-i\hbar k^2t/2m}.
\label{exp}
\end{eqnarray}
Feeding the expression for $z$ appearing in Eq. (\ref{z}) into Eq. (\ref{exp}) allows to write
$\psi^a_n(x,t)$ as
\begin{eqnarray}
\psi^a_n(x,t) &=& ir_n\kappa_ne^{-i\kappa_nL} \times \nonumber \\[.3cm]
&&  De^{ik_0x - i\hbar k_0^2t/2m}  M(y_n^{\prime}),
\label{9}
\end{eqnarray}
where
\begin{equation}
D=-2i(2\pi)^{1/4}\sqrt{\sigma/2},
\label{dz0}
\end{equation}
that follows using $\sqrt{\omega(iz_0)} = \exp(z_0^2/2) \sqrt{{\rm erfc}(z_0)}$ with
$\sqrt{{\rm erfc}(z_0)} \approx 2$ for $z_0 \ll -1$ \cite{faddeyeva, abramowitz},
and $M(y_n^{\prime})$ stands for the Moshinsky function, defined as \cite{gcr97,moshinsky52},
\begin{eqnarray}
M(y_n^{\prime}) &=& \frac{i}{2\pi}\int_{-\infty}^\infty dk \frac{e^{ikx'-i\hbar k^2t'/2m}}{k-\kappa_n'} \nonumber \\ [.3truecm]
&=&\frac{1}{2} e^{imx'^2/2\hbar t'} \omega(iy_n'),
\label{11}
\end{eqnarray}
with
\begin{equation}
x'= x - x_c -\frac{\hbar k_0}{m}t, \quad t' = t-i\tau, \quad \kappa_n'= \kappa_n-k_0,
\label{10}
\end{equation}
and the argument $y_n'$ of the Faddeyeva function $\omega(iy_n')$ reads,
\begin{equation}
y_n'=e^{-i\pi/4} \sqrt{\frac{m}{2\hbar t'}} \left [ x'-\frac{\hbar \kappa_n'}{m}t'\right]
\label{10aa}
\end{equation}
The Moshinsky function is usually calculated via the  Faddeyeva functions for which
well developed computational routines are available \cite{poppe},
Substitution of Eq. (\ref{9}) into Eq. (\ref{3ap}), allows to write the time-dependent transmitted solution as,
\begin{widetext}
\begin{equation}
\psi^a(x,t)= C \psi^a_f(x,t) + i De^{ik_0x - i\hbar k_0^2t/2m}
\sum_{n=-\infty}^{\infty} r_n\kappa_ne^{-i\kappa_nL}  M(y_n^{\prime}).
\label{18}
\end{equation}
\end{widetext}
Notice, in view of the definitions for $x'$ and $t'$ in Eq. (\ref{10}) and of $\tau$ given by Eq. (\ref{tau}), that the argument
of the second exponential term on  the right-hand side of Eq. (\ref{free}) may be written as $-imx'^2/2\hbar t'$. This allows to write
Eq. (\ref{free}) for $\psi^f_a(x,t)$ as
\begin{equation}
\psi^a_f(x,t)=\frac{1}{(2\pi)^{1/4}}\frac{1}{\sigma^{1/2}} \frac{e^{i(k_0x-\hbar k_0^2t/2m)}e^{-imx'^2/2\hbar t'}}{\sqrt{1+it/\tau}}
\label{free2}
\end{equation}
and  Eq. (\ref{18}), using Eq. (\ref{11}), alternatively, in the more convenient form as
\begin{widetext}
\begin{equation}
\psi^a(x,t)=\psi^a_f(x,t)\left [ C +  \pi^{1/2}\sigma \sqrt{1+it/\tau}
\sum_{n=-\infty}^{\infty} r_n\kappa_ne^{-i\kappa_nL}  \omega(y_n^{\prime})\right ].
\label{18f}
\end{equation}
\end{widetext}
One should emphasize that Eqs. (\ref{free2}) and (\ref{18f})  hold provided the condition given
by Eq. (\ref{8bb}) is satisfied.

Clearly, as a consequence of the approximation given by Eq. (\ref{omega}), the solution  $\psi^a(x,t)$ does not
vanish exactly as $t \rightarrow 0$. There remains a small value proportional to the tail of the free solution.

\subsection{Long-time behavior of $\psi^a(x,t)$}

Let us now analyze  Eq. (\ref{18f}) at asymptotically long times, \textit{i.e.} much larger than lifetime $\tau$
of the system,  for a fixed value of the distance $x=x_d$. In such a case, one sees from Eq. (\ref{10aa}), that the argument $y'_n$ of the Faddeyeva functions behaves as
\begin{equation}
y_n' \approx - e^{-i\pi/4} \sqrt{\frac{\hbar}{2m}} \kappa_nt^{1/2}; \quad x=x_d, \quad t \gg \tau,
\label{10aas}
\end{equation}
and hence  becomes very large as time increases. For proper poles $\kappa_n=\alpha_n-i\beta_n$, \textit{i.e.}, $\alpha_n > \beta_n$, the Faddeyeva function behaves as \cite{faddeyeva,abramowitz},
\begin{equation}
w(iy'_n) \approx  e^{y'^2_n} + \frac{1}{\pi^{1/2}}\left( \frac{1}{y'_n} - \frac{3/2}{{y'}_n^3} \right ) + \cdots,
\label{asym1}
\end{equation}
Using Eq. (\ref{10aas}) it follows that the term $\exp(y'^{\,2}_n)$ vanishes exponentially with time. On the other hand,
for poles $\kappa_{-n}=-\kappa_n^*$, seated on the third quadrant of the $k$ plane, the Faddeyeva function behaves in a purely nonexponential fashion as on the right hand-side of Eq. (\ref{asym1}) \cite{faddeyeva,abramowitz}, namely
\begin{equation}
w(iy'_{-n}) \approx \frac{1}{\pi^{1/2}}\left( \frac{1}{y'_{-n}}  -
\frac{3/2}{{y'}_{-n}^3} \right ) + \cdots,
\label{asym2}
\end{equation}
One sees therefore, that for sufficiently long times the full set of resonance poles behaves nonexponentially.
Using Eqs. (\ref{coefc}), (\ref{asym1}) and (\ref{asym2}), one may write Eq. (\ref{18f}) at asymptotically long times as,
\begin{widetext}
\begin{equation}
\psi^a(x,t)\approx \psi^a_f(x,t)\left [ i\sum_{n=-\infty}^{\infty} r_ne^{-i\kappa_nL} +
e^{i \pi/4} t^{1/2}\frac{1}{\sqrt{2m/\hbar}}
\sum_{n=-\infty}^{\infty} r_n\kappa_ne^{-i\kappa_nL}\left ( \frac{1}{y'_n} - \frac{3/2}{{y'}_n^3} \right ) \right ].
\label{18asym}
\end{equation}
\end{widetext}
Substitution of Eq. (\ref{10aas}) into Eq. (\ref{18asym}), one sees that the second term on the right-hand side cancels exactly the first one and hence one obtains that $\psi^a(x,t)$ behaves as,
\begin{equation}
\psi^a(x,t) \sim\frac{1}{t^{3/2}}; \quad x=x_d, \quad  t \gg \tau.
\label{long}
\end{equation}
It follows from the above expression that the corresponding probability density goes as $1/t^3$.
This long-time behavior of the probability density as an inverse cubic power of time has also been obtained
with other potential models and initial states, including numerical calculations of Gaussian wave packets
colliding with square barriers \cite{mdsprb95}. As pointed out in Ref. \cite{mdsprb95} the above
long-time behavior for the probability density is consistent with the definition of the dwell time as a
physical meaningful quantity.

\subsection{Asymptotic behavior of $\psi^a(x,t)/\psi^a_f(x,t)$}

There is another asymptotic limit involving the transmitted wave packet solution given by Eq. (\ref{18f}).
This refers to the limit of  $\psi^a(x,t)/\psi^a_f(x,t)$ as $x \to \infty$ and $t \to \infty$. Previous
analysis regarding the time evolution of forerunners involving  cutoff initial plane waves show  that
at very large distances and long times, $x/t \to (\hbar k/m)$ \cite{vr03}. This suggest a similar behavior
for the transmitted Gaussian pulse.
Hence as $x$ and $t$ attain very large values, one may write the argument $y'_n$ of the Faddeyeva function,
given by Eq. (\ref{10aa}) as
\begin{equation}
y_n' \approx  e^{-i\pi/4} \sqrt{\frac{\hbar}{2m}}[k- \kappa_n]t^{1/2}; \quad x=(\hbar k/m)t, \quad t \to \infty,
\label{10aasy}
\end{equation}
where the relationships given by Eq. (\ref{10}) have been used. It follows then, using the leading $1/y'$ terms
in  Eqs. (\ref{asym1}) and (\ref{asym2}),  that at very long times the term
$\sqrt{1+it/\tau}\,\, \omega(iy'_n)$ appearing in Eq. (\ref{18f}) tends to $i/[(\pi^{1/2}\sigma)(k-\kappa_n)]$. As a
consequence, one may rewrite Eq. (\ref{18f}), for very large values of $x$ and $t$ as
\begin{equation}
\frac{\psi^a(x,t)}{\psi^a_f(x,t)} = {\bf t}(k); \quad x=(\hbar k/m)t, \quad t \gg \tau
\label{19}
\end{equation}
where Eqs. (\ref{coefc}) and (\ref{3}) have been used. Equation (\ref{19}) provides an   analytical demonstration that
at very large  distances and times, $\psi^a(x,t)/\psi^a_f(x,t)$ reproduces  the transmission amplitude of the
system, and hence $|\psi^a(x,t)/\psi^a_f(x,t)|^2=T(E)$ \textit{vs} $E$, the corresponding transmission energy spectra
of the system.

\section{Examples and discussion}

In order to exemplify our findings, we consider three
tunneling systems involving typical parameters of semiconductor
$Al_xGa_{1-x}As$ materials \cite{ferry}. The first one is a single
barrier (SB) with barrier width $b=8$ nm and barrier height
$V=0.23$ eV. The second system is a double-barrier resonant
tunneling structure (DB) with barrier width $b = 5.0$ nm, well
width $w=5.0$ nm and barrier heights $V = 0.23$ eV. The third
system refers to a quadruple-barrier resonant tunneling structure
(QB), with external barrier widths $b_1= b_4 = 3.0$ nm, internal
barrier widths $b_2=b_3=5.0$ nm, well widths $w_1 = w_2=w_3= 3.0$
nm and barrier heights $V=0.23$ eV. In all three systems the effective
electron mass is taken as $m = 0.067\ m_e$ where $m_e$ is the
electron mass.

\begin{figure}[!tbp]
\begin{center}
\includegraphics[width = 8cm]{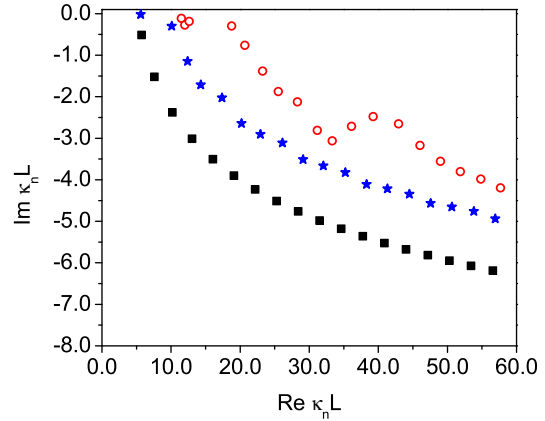}
\caption{ \footnotesize (Color on line) Distribution of the first poles of the outgoing Green function on the $kL$ plane, SB (full squares), DB (full stars) and QB (empty circles), where the parameter $L$ is the length corresponding
to each system. See text.}
\label{f1}
\end{center}
\end{figure}
\begin{figure}[!tbp]
\begin{center}
\includegraphics[width = 7cm]{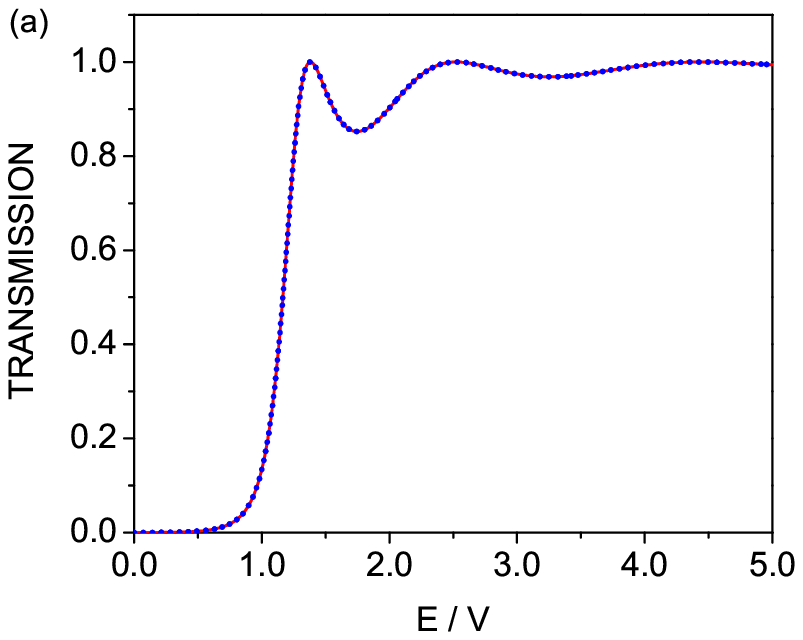}
\includegraphics[width = 7cm]{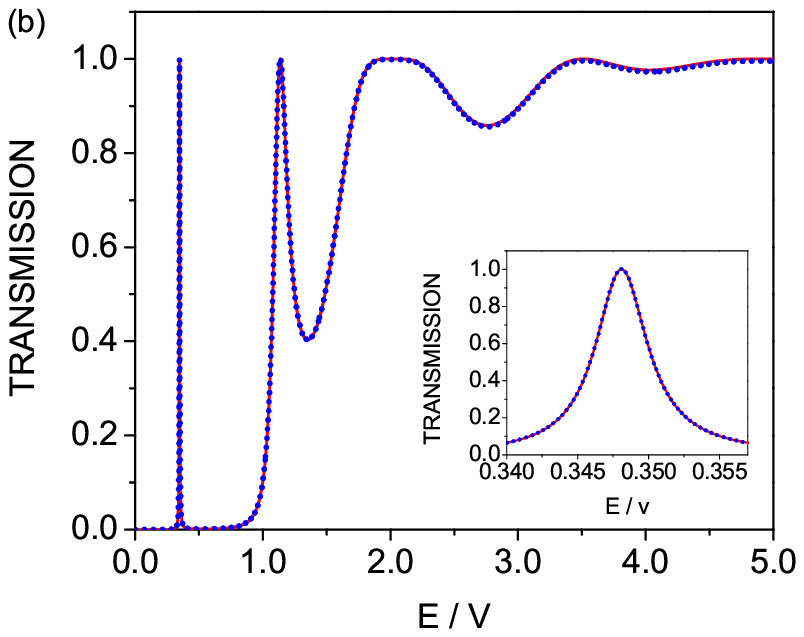}
\includegraphics[width = 7cm]{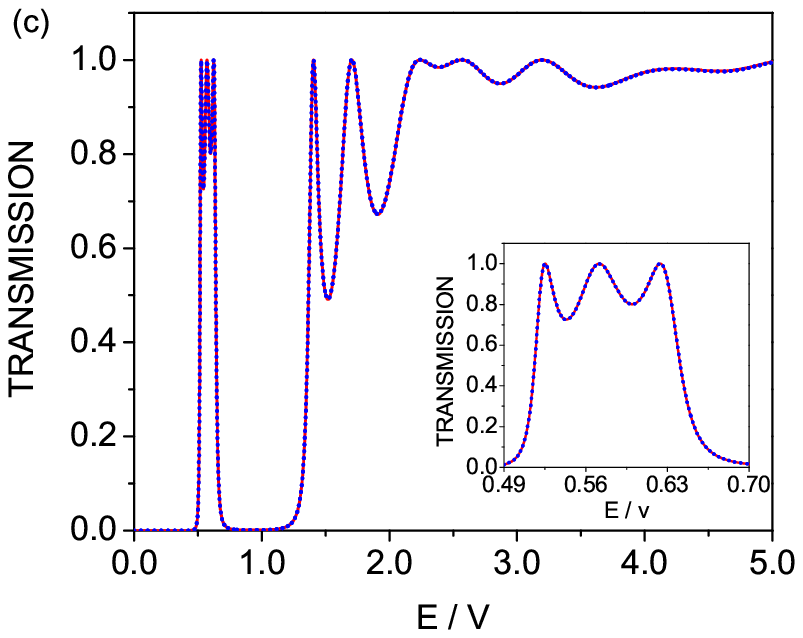}
\caption{ \footnotesize (Color on line) Shows the exact
transmission (full line) in comparison with the analytical formula
(Eq. \ref{3}) (dot line)  for (a) The SB system, using an
approximation of 300 poles of the outgoing Green function;  (b) The DB
system, using an approximation of 1000 poles, and (c) The QB system,
using an approximation of 4000 poles. The inset shows the first
isolated resonance and miniband corresponding at DB and QB systems
respectively. }\label{f2}
\end{center}
\end{figure}

For a given potential profile, the parameters of the system
determine the values of the complex poles $\{\kappa_n\}$  which are the relevant ingredients to calculate the resonance states $u_n(x)$ and hence the residues $r_n$ appearing in both,  Eq. (\ref{3}) for the transmission amplitude, and Eq. (\ref{18f}), for the transmitted time-dependent solution.
Although the procedure to calculate the complex poles is known, for completeness, we present in
Appendix B, a procedure to obtain the necessary number of complex poles involving the Newton-Raphson
method \cite{raphson}. The set of resonance states $\{u_n(x)\}$ may be obtained using the transfer matrix method \cite{ferry} with the outgoing boundary conditions given by Eq. (\ref{bc}).

It is of interest to stress that  a given potential profile provides a unique set of resonance poles $\{\kappa_n\}$ and residues $\{r_n\}$ that are calculated only once to evaluate Eq. (\ref{18f}). This implies that calculations are much less time demanding than calculations involving  numerical integration of the solution given by Eq. (\ref{1}) where one has to perform an integration over $k$ at each instant of time, particularly if one is interested, as in the present work, to evaluate the above solution at very long times and distances.

\subsection{Complex poles and Transmission coefficient}

Figure \ref{f1} exhibits the distribution of the first complex
poles for the SB (full squares), DB (full stars) and the QB
(empty circles) systems with parameters as given above. In order to
facilitate a comparison among the distinct distributions, the complex poles
of each system are multiplied by the corresponding total length
$L$, \textit{i.e.}, respectively, for the SB, DB and QB systems:
$L=8.0$ nm, $L=15.0$ nm  and $L=25.0$ nm.

Figures \ref{f2}(a), \ref{f2}(b), and \ref{f2}(c), show respectively,
for the SB, DB and QB systems considered above, a plot of the transmission coefficient
$T(E) = |{\bf t}(E)|^2$ as function of the energy $E$ in units of
the barrier height $V$. Each figure presents a comparison between an exact
numerical calculation using the transfer matrix method (full line),
and that obtained using the resonance expansion given by Eq. (\ref{3}) (dotted line).
One observes that the calculations are indistinguishable from each
other if one considers the appropriate number of poles, as indicated for each case in the caption to Fig. \ref{f2}.

Let us comment briefly some features of each of the above figures.

Figure \ref{f2}(a) exhibits a broad overlapping resonance just above the barrier height that is related
to the presence of the complex energy pole, $E_1 = \mathcal{E}_1-i\Gamma_1/2$,  with values $\mathcal{E}_1 = 0.2885$ eV
and $\Gamma_1 = 0.1045$ eV.

Notice in Figure \ref{f2}(b), describing a DB system,  the existence of a sharp isolated resonance in
the tunneling region with a typical Breit-Wigner or Lorentzian
shape as exhibited by the inset. The corresponding
resonance energy parameters are: $\mathcal{E}_1=0.08$ eV and $\Gamma_1=1.0278$ meV. At
energies above the barrier height the DB system exhibits some transmission resonance structures that
tend to disappear as the energy increases.

Figure \ref{f2}(c), involving the QB system,  exhibits a triplet of overlapping resonances along the tunneling region,
displayed enlarged in the inset. The corresponding resonance energies are,
respectively, $\mathcal{E}_1=0.1199$ eV, $\mathcal{E}_2=0.1309$ eV
and $\mathcal{E}_3=0.1450$ eV and the corresponding widths,
$\Gamma_1=4.6270$ meV, $\Gamma_2=11.9652$ meV and
$\Gamma_3=8.4472$ meV. Similarly, as in the case of the DB system,
the QB system exhibits transmission resonances above the barrier
height as the energy increases. Notice that the triplet of
overlapping resonances corresponds to the first triplet of
resonance poles exhibited in Fig. \ref{f1} (empty circles). This
triplet of resonance poles suffices to reproduce the transmission
coefficient around the corresponding  energy range \cite{gcrr93}.

\subsection{Time evolution of the transmitted probability density}
\begin{figure}[!tbp]
\begin{center}
\includegraphics[width = 7.7cm]{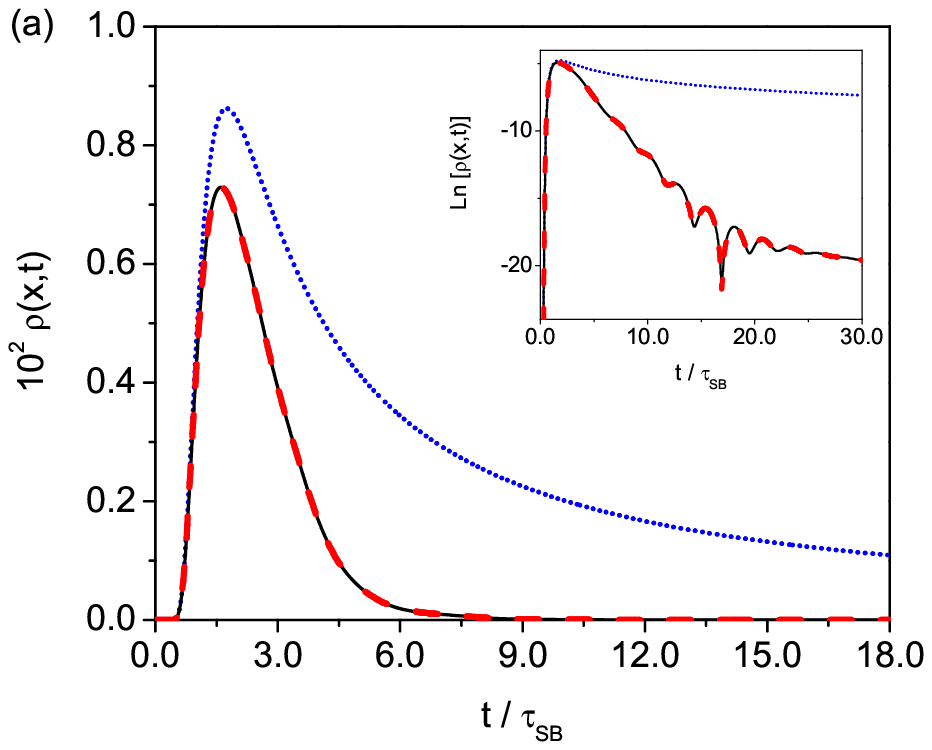}\\  %
\includegraphics[width = 7cm]{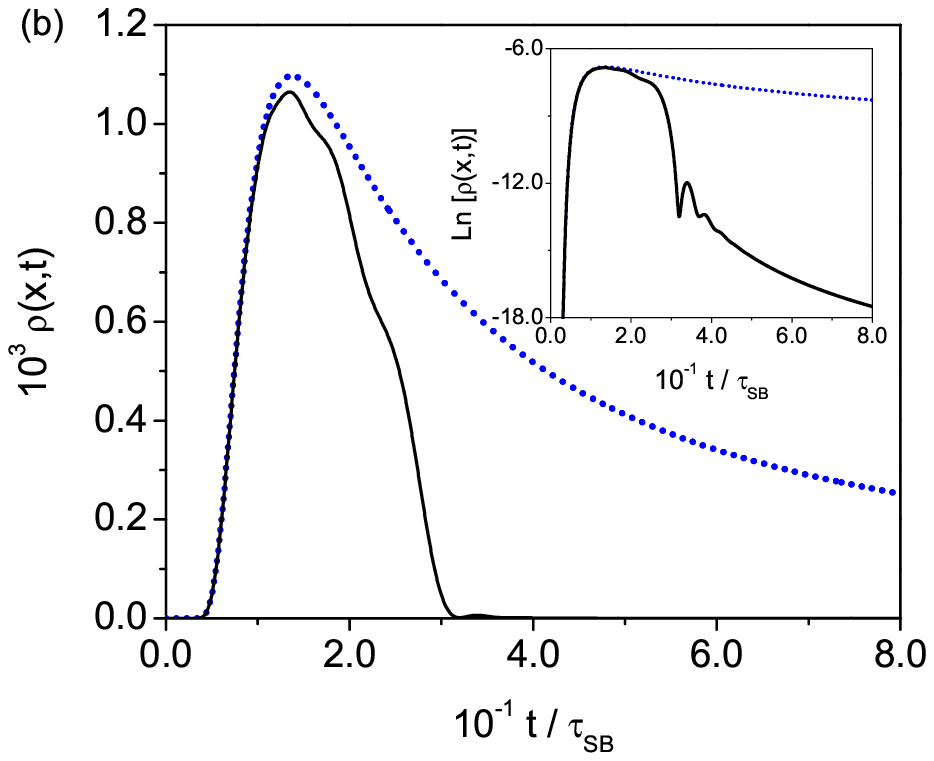}\\  %
\includegraphics[width = 7.7cm]{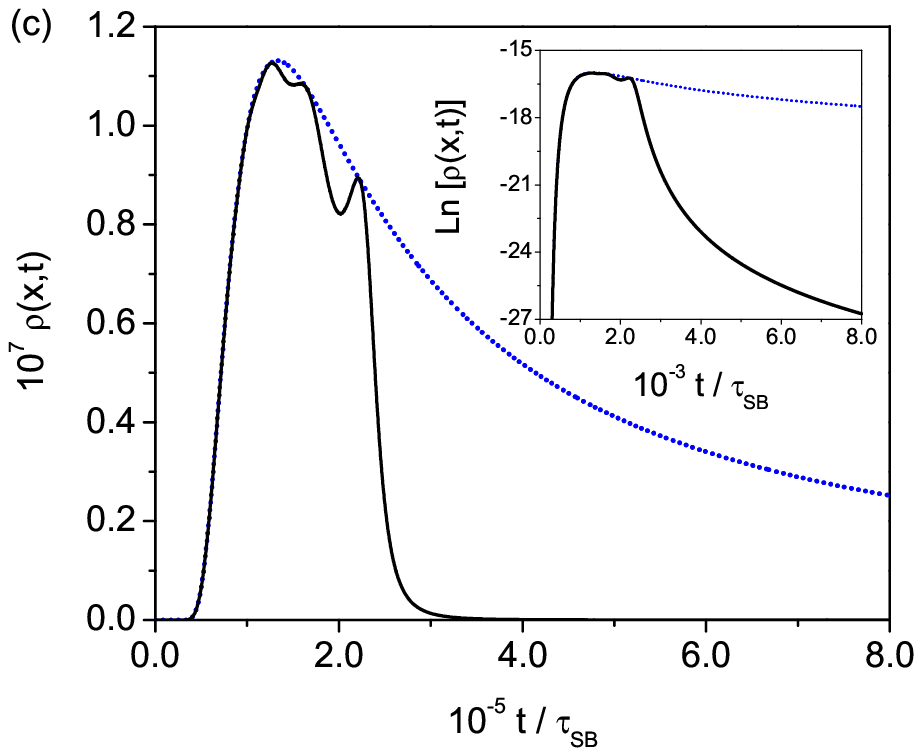}  %
\caption{ \footnotesize (Color on line) Probability density as function of the time, in units of $\tau_{SB}=\hbar/\Gamma_1$,
for the SB system (full line). The energy of the initial Gaussian state is  $E_0=V/2$.
As a comparison, the free Gaussian evolution  (dotted line) is also plotted. The calculation is performed at  (a)
$x_d=2L$ short, (b) $x_d = 20L$ medium and (c) $x_d=2\times10^5L$ long, distances,  where $L$ is the length of the SB system. The inset shows the calculation in a semi-ln scale. The exact calculation by numerical integration using
Eq. (\ref{1}) is displayed also in (a) (dashed line). See text.}\label{f3}
\end{center}
\end{figure}
Let us now investigate the time evolution of the transmitted probability density $|\psi(x,t)|^2$ using Eq. (\ref{18f}) as time evolves for different values of $x=x_d$.
We find convenient to plot the dimensionless quantity $\rho(x,t)=\sigma|\psi(x,t)|^2$ in units of $t/\tau$,
where $\tau$  stands for the longest  lifetime of the system, \textit{i.e.}, $\tau \equiv \hbar/\Gamma_{min}$,
with $\Gamma_{min}$ the smallest energy width.

The parameters of the initial cutoff Gaussian wave packet, defined  by Eq. (\ref{5}), are
\begin{equation}
x_c =-5.0\, {\rm nm}, \hskip1truecm \sigma = 0.5\,{\rm nm}.
\label{pargaus}
\end{equation}
These values  give  $|x_c|/(2\sigma)=5.0$, which implies that the condition
given by Eq. (\ref{8bb}) is satisfied, and hence the applicability of Eq. (\ref{18f}), to calculate the time evolution
of the transmitted probability density. Notice that $\sigma < L$ for all the systems considered, \textit{i.e.},
$L_{SB}=8.0$ nm, $L_{DB}=15.0$ nm  and $L_{QB}=25.0$ nm.
Also, we choose $E_0=V/2$ for the SB, $E_0=\mathcal{E}_1$ for the DB system and  $E_0=\mathcal{E}_2$
for the QB system. The values of the natural time scale  $\tau$ are  $\tau_{SB} = 5.57$ fs for the SB
system, $\tau_{DB} =0.64$ ps for the DB system, and $\tau_{QB}=0.14$ ps for the QB system.

\begin{figure}[!tbp]
\begin{center}
\includegraphics[width = 7.7cm]{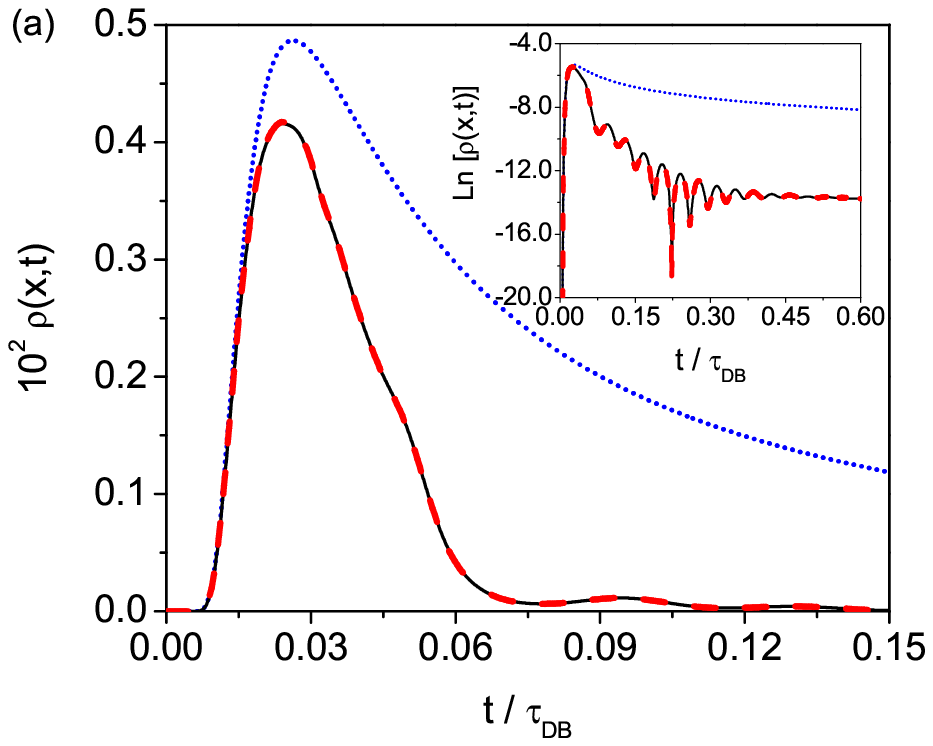}\\  %
\includegraphics[width = 7cm]{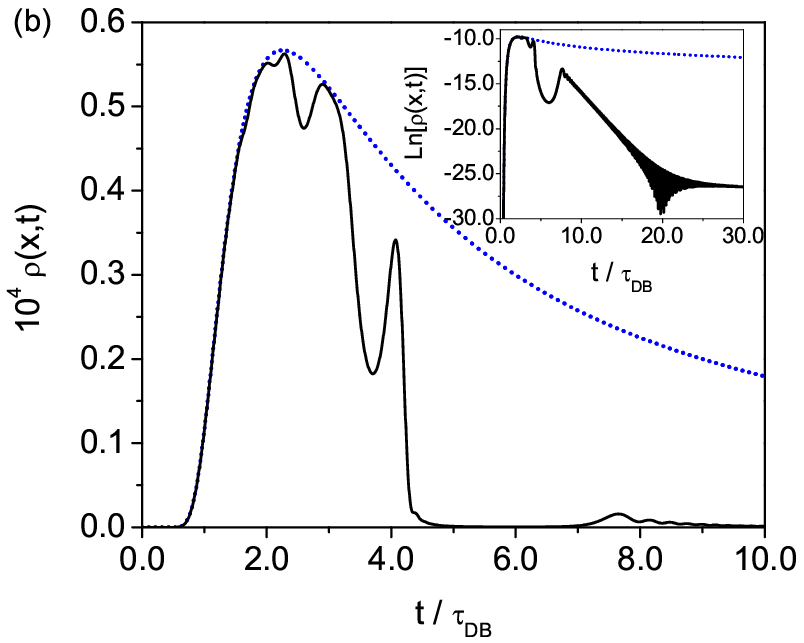}\\  %
\includegraphics[width = 7cm]{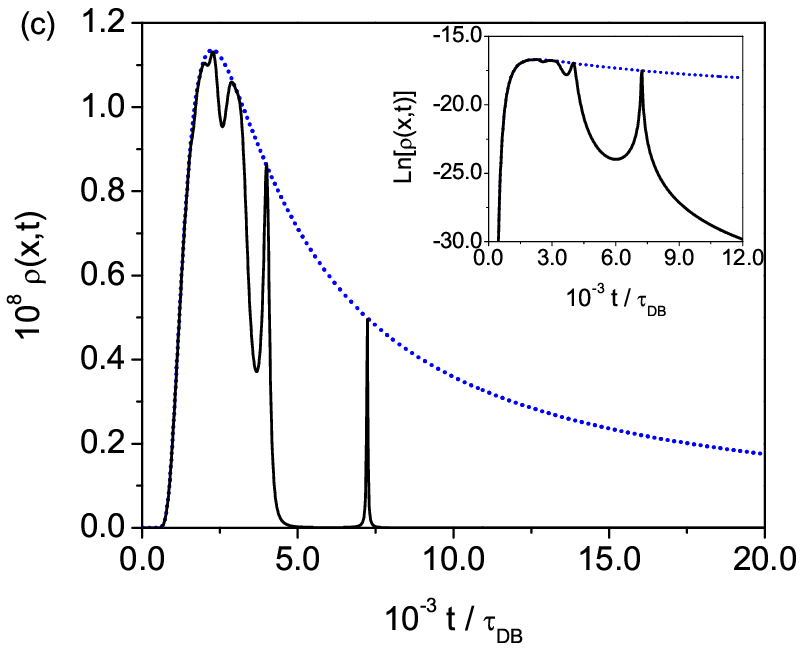}  %
\caption{ \footnotesize (Color on line) Probability density
as function of the time, in units of $\tau_{DB}=\hbar/\Gamma_1$,
for the DB system (full line). The energy of the initial Gaussian state corresponds to the resonance energy, \textit{i.e.},
$E_0=\mathcal{E}_1$. As a comparison, the free Gaussian wave packet (dotted line) is plotted. The calculation
is performed at  (a) $x_d=2L$ short, (b) $x_d = 200L$ medium and (c)
$x_d=2\times10^5L$ long,  distances. The parameter $L$ stands for the length of the DB
system. The  inset shows the same calculation in a semi-ln scale. The exact calculation by numerical integration using Eq. (\ref{1}) is
displayed also in (a) (dashed line). See text.}\label{f4}
\end{center}
\end{figure}

Figure \ref{f3} shows a plot of $\rho(x_d,t)$ \textit{vs} $t/\tau$ for the SB system (full line) for different values of $x_d$ that correspond to short, medium and long distances from the interaction region, (a)  $x_d = 2L$, (b) $x_d=20L$, and (c) $x_d = 2\times10^5L$. The results are compared with the corresponding free propagation of the cutoff Gaussian wave packet (dotted line).
The distinct distances represent different time scales in the time evolution of the transmitted wave packet.
Notice that in all the three cases at short times, the profiles of the free and transmitted wave packets are essentially the same. This follows by noticing that the large  over-the-barrier energy components of the wave packet that impinge on the potential barrier are transmitted without suffering an appreciable change, as exhibited by the behavior of the corresponding  transmission coefficient displayed in Fig. \ref{f2} (a) ($T(E) \to 1$ as $E/V \gg 1$), and by using a Fourier transform argument that indicates that short times correspond to large energies. Figure \ref{3}(a) shows that after reaching its maximum value the transmitted wave packet (solid line) decays faster than the free evolving wave packet (dotted line).  As shown by the inset to Fig. \ref{f3}(a), this is so because that time span is dominated by the exponential decay of the first top resonance, which in fact after a number of lifetimes suffers a transition into a nonexponential behavior as an inverse $t^{-3}$  power of time as follows from  Eq. (\ref{long}).
\begin{figure}[!tbp]
\begin{center}
\includegraphics[width = 7.7cm]{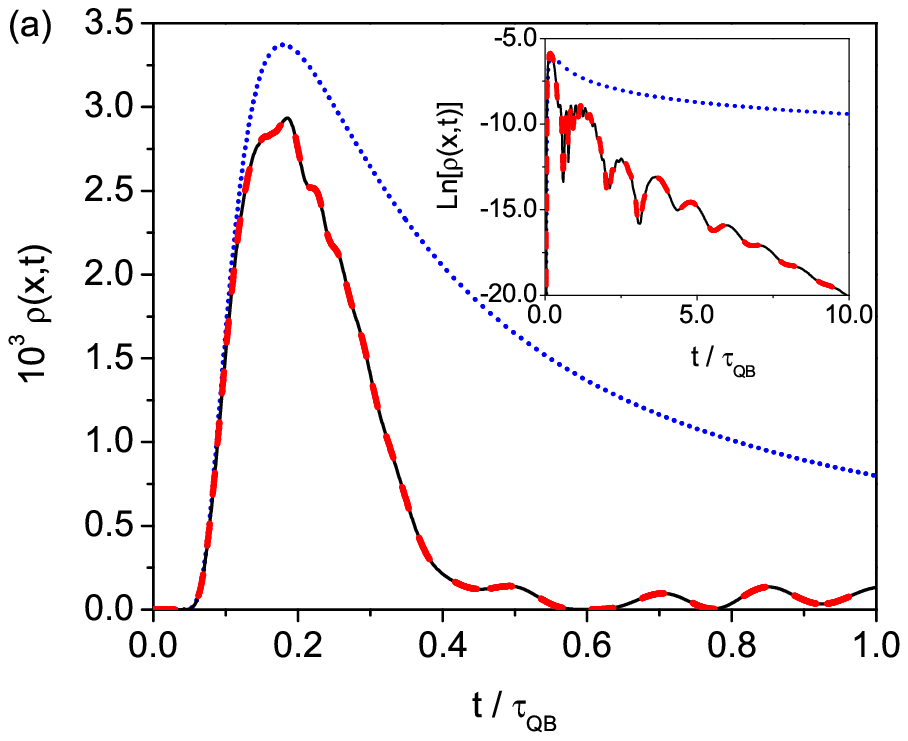}\\  %
\includegraphics[width = 7cm]{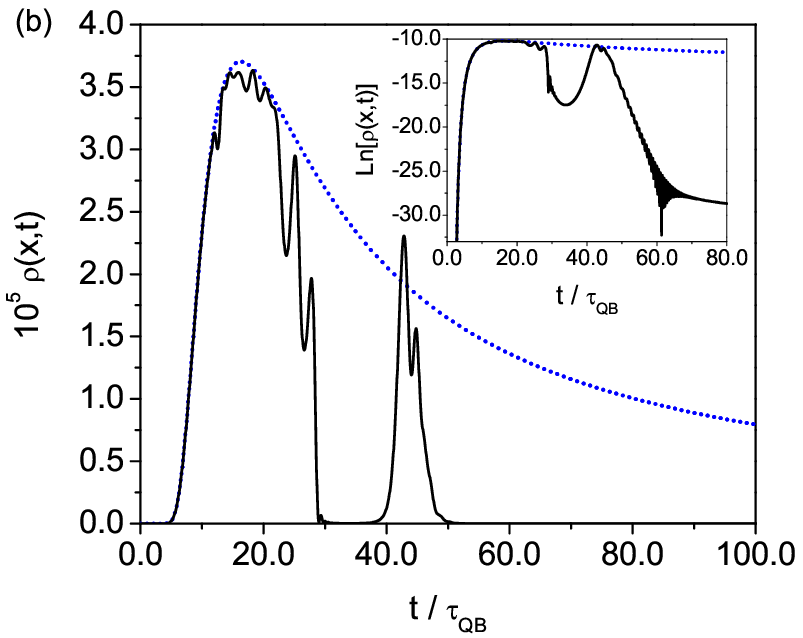}\\  %
\includegraphics[width = 7cm]{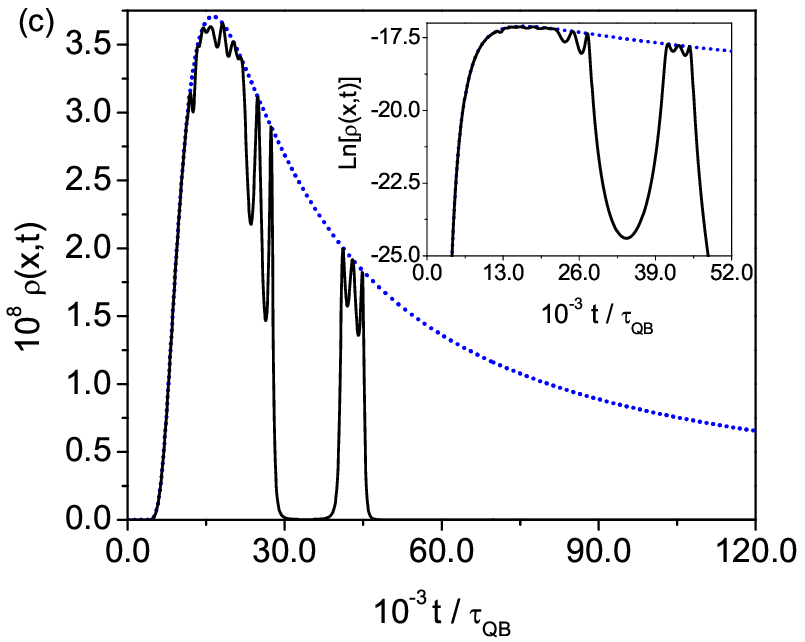}  %
\caption{ \footnotesize (Color on line) Probability density
as function of the time, in units of the lifetime $\tau_{QB} =
\hbar/\Gamma_1$, for the QB system
(full line). The energy of the initial Gaussian state is equal to the resonance energy
$E_0=\mathcal{E}_2$. As a comparison the free  Gaussian wave packet is plotted (dotted line).
The calculation is performed at  (a) $x_d=2L$ short, (b) $x_d = 200L$
medium and (c) $x_d=2\times10^5L$ long distances, where $L$ is the
length of the QB system. The inset shows similar calculations in semi-ln scale.
The exact calculation by numerical integration using Eq. (\ref{1}) is
displayed also in (a) (dashed line). }\label{f5}
\end{center}
\end{figure}
Figures \ref{3}(b) and \ref{3}(c) exhibit the time evolution of the probability density, respectively,
at  medium, $x_d=200L$, and large, $x_d=2 \times 10^5L$, distances and hence medium and long times. The corresponding
probability density profiles are very similar in both figures. The corresponding insets  are also similar and show
that the time evolution goes as the inverse power $t^{-3}$ . At larger distances the profile exhibited by Fig. \ref{3} (c) remains unchanged. One sees that the profile reflects the energy spectra of the system as shown by a comparison with Fig. \ref{2} (a) for the transmission coefficient \textit{vs} energy.

Figure \ref{f4} exhibits an analogous calculation of the transmitted probability density for  the DB system (full line) and its comparison with the free evolving wave packet (dotted line). In this case, at short distances, $x_d=2L$, the inset shows that the profile of the transmitted wave packet is dominated by the transition from the first top resonance (See Fig. \ref{2}(b)) into the sharp isolated resonance seated inside the system. At medium distances, $x_d=200L$, one observes a small peak structure around $t/\tau_{DB} \approx 8$. As the inset displays, it corresponds to the exponential decay of the sharp isolated resonance situated inside the system. This situation is similar to that discussed by Wulf and Skalozub, who considered the propagation of a Gaussian pulse near a resonance level \cite{wulf}. The inset shows that eventually at longer times there is a transition to nonexponential decay as an inverse $t^{-3}$ power of time. Finally at very large distances, $x_d=2 \times 10^5L$ and very long times, of the order of $10^3 \tau_{DB}$, in a similar fashion as in the previous system, the profile of the transmitted wave packet reflects already the structure of the energy spectra of the DB system. The corresponding inset to Fig. \ref{f4}(c) shows that the sharp structure around $t/\tau_{DB} \approx 7.2 \times 10^3$ evolves at long times in a nonexponential fashion.

Figure \ref{f5} exhibits also a situation similar to the examples discussed above, for the time evolution of the transmitted probability density of the QB system. Again Figs. \ref{f5} (a), (b) and (c) refer, respectively, to short, $x_d=2L$, medium $x_d=200L$, and large, $x_d=2 \times 10^5L$, distances. At short distances, $x_d=2L$, it  is worthwhile to notice the presence of \textit{Rabi} oscillations in a similar fashion as occur in the decay of multibarrier systems \cite{gcrv07}. These oscillations represent transitions among the closely lying resonance levels of the QB system. Again as the distance and the time increase, the resonance levels decay, first exponentially and then nonexponentially, as depicted in the inset to Fig. \ref{f5}(b). At still larger distances the decay is purely nonexponential, as an inverse cubic power of time, as shown by the inset to Fig. \ref{f5}(c). Notice that in Fig. \ref{f5}(c) the profile of the transmitted wave packet resembles already the energy structure of the transmission coefficient.

It is of interest to compare our results with the case of a cutoff incident plane wave impinging on a multibarrier system. This case, corresponding  to the limit of an infinitely broad Gaussian wave packet, has been considered recently by Villavicencio and Romo \cite{vr03}, using the  formalism developed in Ref. \cite{gcr97}, to investigate the propagation of transmitted quantum waves in these systems. There, for incidence energies $E_0$ below the lowest resonance energy of the multibarrier system, a series of propagating pulses (forerunners) are observed in the transmitted solution  traveling faster than the main wavefront. It is shown that each forerunner propagates with speed $v(\mathcal{E}_n)$ =$[2m\mathcal{E}_n/m]^{1/2}$ associated with the $n$th resonance  of the system, thus establishing a relationship between the sequence of forerunners and the resonance spectrum of the system. However at asymptotically long times the forerunners fade away, since the solution $\psi(x,t) \sim t(k_0) \exp(ik_0x)\exp(-iE_0t)/\hbar)$, with $k_0=[2mE_0]^{1/2}/\hbar$. This yields for the transmitted probability density $|\psi(x,t)|^2=|t(k_0)|^2$, a result very different from the case of  Gaussian wave packets of finite width considered here.

\begin{figure}[!tbp]
\begin{center}
\includegraphics[width = 8cm]{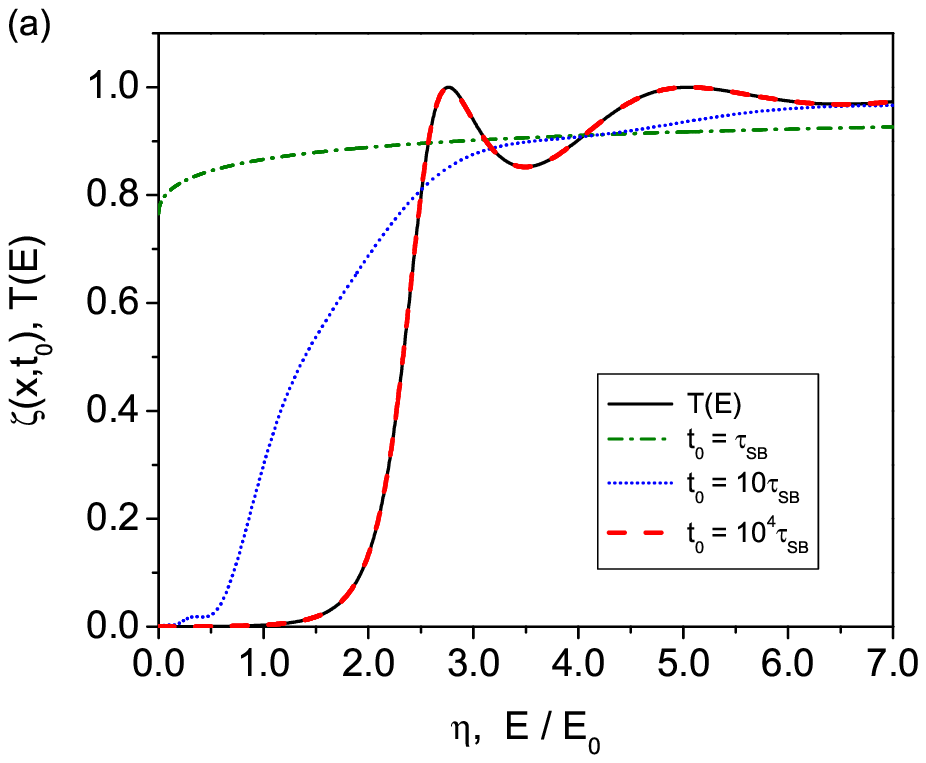}\\  %
\includegraphics[width = 8cm]{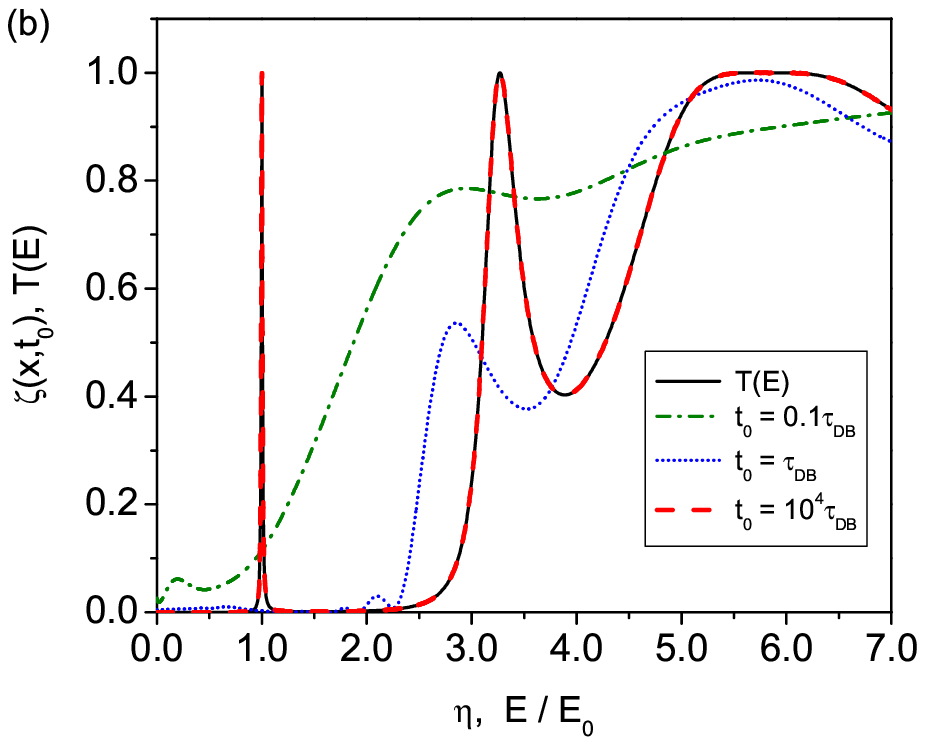}\\  %
\includegraphics[width = 8cm]{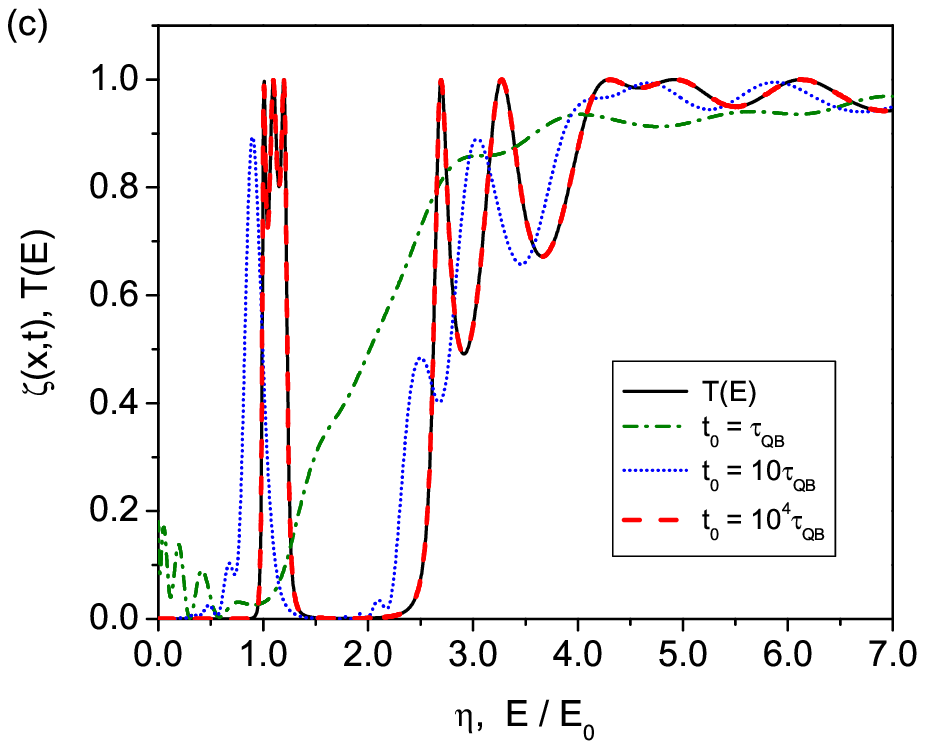}  %
\caption{ \footnotesize (Color on line) Figures (a), (b), and (c) refer, respectively, to the single barrier (SB),
double-barrier (DB) and quadruple-barrier (QB) systems. In each figure:  $T(E)$ stands for the transmission
coefficient in units of $E_0$, the energy of the incident wavepaket (solid line); $\zeta(x,t_0)$ defined by
Eq. (\ref{newT}) in units of the parameter $\eta$, for distinct values of $t_0$,
as specified in the insets to each figure. As time increases the transients end by reproducing the energy spectra of the
corresponding systems. See text.}
\label{f6}
\end{center}
\end{figure}
\subsection{Reconstruction of the energy spectra}

In order to exhibit more clearly the relationship between the time evolution of the  transmitted Gaussian wave packet
and the energy spectra of the system, pointed out in the previous subsection, one may proceed as follows.
First, instead of evaluating the transmitted probability density by fixing $x=x_d$ and varying the time $t$, \textit{i.e.}, $|\psi(x_d,t)|^2$,
as discussed in the previous subsection, we consider instead  a fixed value of the time $t=t_0$ and vary $x$, \textit{i.e.}, $|\psi(x,t_0)|^2$.
It is not difficult to see that the plot of $|\psi(x,t_0)|^2$ \textit{vs} $x$ looks identical to the specular image, with respect to the vertical axis at the origin, of $|\psi(x_d,t)|^2$ \textit{vs} $t$.
Second, in analogy with the calculation of the transmission coefficient in the energy domain, we divide
$|\psi(x,t_0)|^2$ by the free evolving Gaussian wave packet $|\psi_a^f(x,t_0)|^2$, given by Eq. (\ref{free}).
We define the quantity $\zeta(x,t_0)$ as the ratio of these quantities, namely,
\begin{equation}
\zeta(x,t_0)= \frac{|\psi(x,t_0)|^2}{|\psi_a^f(x,t_0)|^2}.
\label{newT}
\end{equation}
Third, it is convenient to plot the transmission coefficient $T(E)$ in units of $E/E_0$, with $E_0$, the incident energy of the corresponding Gaussian wave packet. This allows to relate the values of  $E/E_0$ with the values of
a parameter $\eta$ defined as
\begin{equation}
\eta \equiv  \left [\frac{x-L}{x_0-L} \right ]^2,
\label{eta}
\end{equation}
as follows. The above expression for $\eta$  is based on the argument that for $x \gg L$,
$E=\hbar^2 k^2/2m$ with $\hbar k/m=(x-L)/t_0$, and $E_0=\hbar^2 k_0^2/2m$ with $\hbar k_0/m=(x_0-L)/t_0$,
where $x_0-L$ is the distance that a free particle travels in time $t_0$. Hence, $E=\eta E_0$.
Figures \ref{f6} (a), (b) and (c) exhibit respectively, for the SB, DB, and QB
systems, considered in the previous subsection, the plots of $\zeta(x,t_0)$ \textit{vs} $\eta$. For each of the above
figures, three graphs are plotted. As indicated in the inset to each figure, each graph of $\zeta(x,t_0)$ corresponds
to a distinct value of $t_0$ and hence of $x_0-L$. The above figures also exhibit a plot of $T(E)$ in units of
$E/E_0$ (solid line).
Notice, as pointed out previously, that the value of $E_0$ differs for each system.
One may appreciate, in each figure, the transient behavior of the transmitted Gaussian wave packet.
For small values of $t_0$, $\zeta(x,t_0)$  just reproduces the fastest components of the energy spectra
of the corresponding system and as $t_0$ increases it goes into a transient behavior
that ends when the transmission energy spectra of the system is reconstructed, as shown analytically
by Eq. (\ref{19}).

\subsection{Remark on the tunneling time problem}

Our results are of relevance for the tunneling time problem \cite{landauer}. Here, the question posed is:
How long it takes to a particle to traverse a classical forbidden region? One of the approaches considered
involves the tunneling of  wave packets.  Here one usually compares some feature of the incident free evolving
wave packet (usually a Gaussian wave packet) and a comparable feature of the transmitted wave packet, commonly the peak
or the centroid, and a delay is calculated. Many years ago, B\"uttiker and Landauer \cite{buttiker} argued that
such a procedure seems to have little physical justification because an incoming peak or centroid does not, in any
obvious causative sense, turn into an outgoing peak or centroid, particulary in the case of strong deformation
of the transmitted wave packet. Our results for the transient behavior of the transmitted wave packet supports that
view independently of whether or not there is initially a strong deformation of the transmitted wave packet.
Even if the transmitted wave packet is initially no deformed, as in Fig. \ref{f3} (a) for a single barrier system,
as time evolves the profile of the transmitted Gaussian wave packet  varies to finally reproduce  the energy spectra
of the system and hence there is no a unique way to answer the question of how long it took to the initial packet to
traverse the system.

\section{Concluding remarks}

The main result of this work is given by Eq. (\ref{18f}), which provides an analytical solution to the time evolution of a Gaussian wave packet along the transmission region for scattering by a finite range potential in one dimension. We have focused our investigation to cases where the  Gaussian wave packet is initially far from the interaction region, \textit{i.e.} fulfills Eq. (\ref{8bb}), and is sufficiently broad in momentum space so that all sharp and broad resonances of the system are included in the dynamical description. We have obtained analytically and exemplified numerically for single and multibarrier quantum systems, that the profile of the transmitted Gaussian wave packet, exhibits a transient behavior that at large distances and long times becomes proportional to the transmission amplitude of the system, \textit{i.e.}, Eq.(\ref{19}).  This predicts the final destiny of the transmitted wave packet in a coherent process. It is also worth to emphasize that the analytical expression for the transmitted wave packet yields, at a fixed distance and asymptotically long times, a $t^{-3/2}$ behavior with time, \textit{i.e.}, Eq. (\ref{long}). This result corroborates numerical calculations for Gaussian wave packets colliding with square barriers and extends previous analysis to arbitrary potentials of finite range \cite{mdsprb95}. One should recall that the set of poles $\{\kappa_n\}$ and residues $\{r_n\}$, which is unique for a given potential profile, is evaluated just once to calculate the time-dependent solution given by Eq. (\ref{18f}). The number of poles required for a dynamical calculation corresponds to the number of poles necessary  to reproduce the exact transmission amplitude using Eq. (\ref{3}). This is in contrast with calculations involving numerical integration of the solution, using Eq. (\ref{1}), where one has to perform an integration over $k$ at each instant of time and hence the calculation is much more demanding computationally particularly at large distances and long times. Further work is required to extend the results of the present investigation to wave packet dynamics in multidimensional tunneling \cite{mark}. Our analytical solution for the transmitted wave packet might be of interest in connection with the long debated tunneling time problem.

\section{acknowledments} We would like to thank R. Romo for illuminating discussions and the partial
financial support of DGAPA-UNAM IN115108.
\appendix
\section{analysis of $\omega(-iz)$.}

Here we show that the contribution of the term $\omega(-iz)$, appearing on the right-hand side of Eq. (\ref{8aba}),
to the time evolution of the transmitted Gaussian wave packet may be neglected provided the condition given
by Eq. (\ref{8bb})  is fulfilled. The  contribution corresponding to $\omega(-iz)$ reads,
\begin{equation}\label{1b}
\psi^{ne}_n(x,t) = -
\frac{(2\pi)^{1/4}\sqrt{\sigma}}{\sqrt{\omega(iz_0)}} \frac{1}{2\pi}
\int_{-\infty}^\infty \textrm{d} k \frac{\omega(-iz)}{k-\kappa_n}
e^{ikx-i\hbar k^2t/2m},
\end{equation}
where $z$, defined by Eq. (\ref{z}), is written as $z = i(k-k_0')\sigma$ with $k_0' = k_0 - ix_c/2\sigma^2$. In
general, it is necessary to calculate numerically the integral term given by Eq. (\ref{1b}).
However, for  the particular case specified by Eq. (\ref{8bb}), \textit{i.e.},
$|x_c/2\sigma| \gg 1$, that implies that  $|z|>1$ for all values of $k$, one may use the asymptotic expansion of
the Faddeyeva function $\omega(-iz)$ \cite{faddeyeva,abramowitz}
\begin{eqnarray}\label{2b}
\omega(-iz) &\approx& -\frac{i}{\pi} \sum_{j=0}^N
\frac{\Gamma(j+1/2)}{[(k-k_0')\sigma]^{2j+1}} = \nonumber \\ &=&
-\frac{i}{\pi} \sum_{j=0}^N
\frac{\Gamma(j+1/2)}{(2j)!\sigma^{2j+1}}\ D_{k_0'}^{2j}
\frac{1}{k-k_0'},
\end{eqnarray}
where the quantities $D_{k_0'}^{2j}$ denote a $2j$-th derivative operator.

Substitution of Eq. (\ref{2b}) into Eq. (\ref{1b}) allows to express each
integral term in the sum as
\begin{equation} \label{3b}
\frac{i}{2\pi}\int \textrm{d}k \frac{e^{ikx-i\hbar
k^2t/2m}}{(k-k_0')(k-\kappa_n)} = \frac{M(y_0') -
M(y_n)}{k_0'-\kappa_n},
\end{equation}
where we have used the identity
\begin{equation}\label{4b}
\frac{1}{(k-k_0')(k-\kappa_n)} = \frac{1}{k_0'-\kappa_n}\bigg[
\frac{1}{k-k_0'} - \frac{1}{k-\kappa_n}\bigg],
\end{equation}
and the arguments of the  Moshinsky functions $M(y_0')$ and $M(y_n)$
are given respectively by,
\begin{equation}
y_0'=e^{-i\pi/4} \sqrt{\frac{m}{2\hbar t}} \left [ x-\frac{\hbar k_0'}{m}t\right],
\label{10aap}
\end{equation}
and
\begin{equation}
y_n=e^{-i\pi/4} \sqrt{\frac{m}{2\hbar t}} \left [ x-\frac{\hbar \kappa_n}{m}t\right].
\label{10abp}
\end{equation}
Then, the nonexponential contribution of each pole $\kappa_n$ in Eq. (\ref{1b}) reads
\begin{eqnarray}\label{5b}
\psi_n^{ne} (x,t) &=& \frac{(2\pi)^{1/4}\sqrt{\sigma}}
{\sqrt{\omega(iz_0)}} \frac{1}{\pi} \sum_{j=0}^N
\frac{\Gamma(j+1/2)}{(2j)!\sigma^{2j+1}} \times \nonumber \\
&\times& D_{k_0'}^{2j}\bigg[ \frac{M(y_0') -
M(y_n)}{k_0'-\kappa_n} \bigg].
\end{eqnarray}
The dominant term in powers of $\sigma$ in Eq. (\ref{5b}) occurs for $j=0$ and hence the
nonexponential contribution of  each pole is given by
\begin{equation}\label{6b}
\psi_n^{ne}(x,t) \approx \bigg(\frac{2}{\pi}\bigg)^{1/4}\frac{1}{\sqrt{\sigma}}
\frac{M(y_0')-M(y_n)}{(k_0'-\kappa_n)\sqrt{\omega(iz_0)}}.
\end{equation}
Recalling that the factor $\sqrt{\omega(iz_0)} = \exp(z_0^2/2) \sqrt{{\rm erfc}(z_0)}$
and that $z_0 \ll -1$ \cite{faddeyeva, abramowitz} one obtains,
\begin{equation}
\sqrt{\omega(iz_0)} \approx \sqrt{2} e^{x_c^2/4\sigma^2}.
\label{factor}
\end{equation}
It follows then, by substitution of (\ref{factor}) into (\ref{6b}) and comparing the resulting expression
with Eq. (\ref{9}),
taking into account that the corresponding Moshinsky functions yield contributions of the same order of magnitude,
that
\begin{equation}\label{7b}
|\psi_n^{ne}(x,t)| \sim e^{-x_c^2/4\sigma^2}|\psi_n^{a}(x,t)|.
\end{equation}
The above expression demonstrates that provided  Eq. (\ref{8bb}) is satisfied,
the nonexponential contribution $\psi_n^{ne}$  may be neglected.

\section{Calculation of complex poles of the transmission amplitude.}\label{poles}

It is well known that the transmission amplitude $\mathbf{t}(k)$ for a potential $V(x)$ of finite range,
\textit{i.e.}, extending from $x=0$ to $x=L$, possesses an infinite number of complex poles $\kappa_n$ that
in general are simple \cite{newton}.  These complex poles correspond to the zeros of the element $t_{22}(k)$ of the corresponding transfer matrix
\begin{equation}\label{1p}
\mathbf{t}(k) = \frac{1}{t_{22}(k)}.
\end{equation}
\begin{figure}
\begin{center}
\includegraphics[width = 6.5cm]{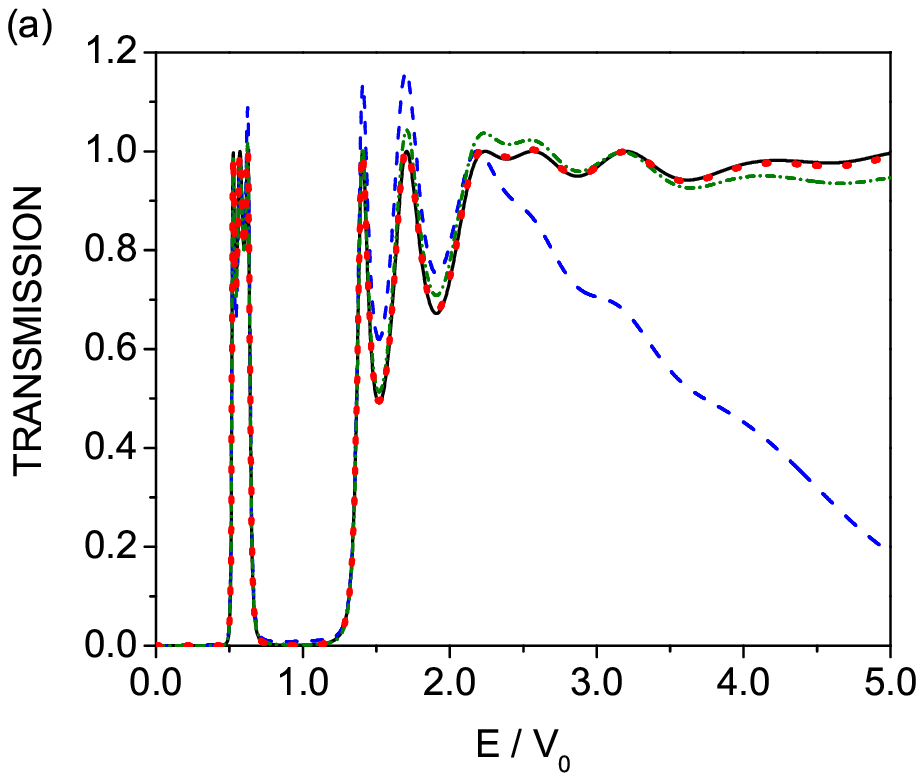} %
\includegraphics[width = 6.5cm]{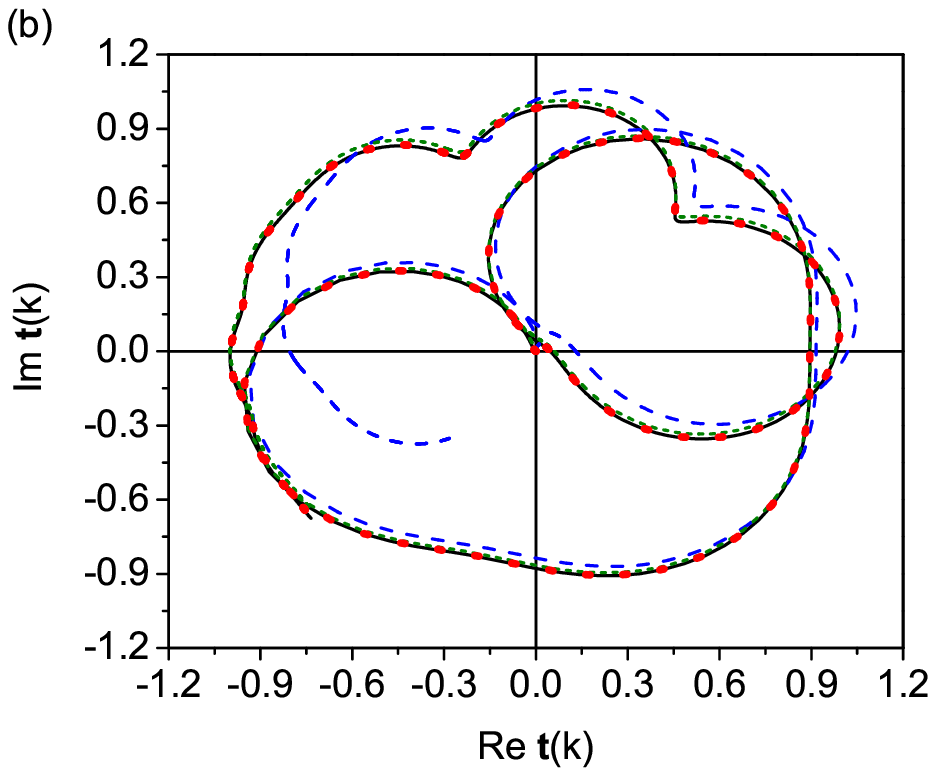} %
\caption{\footnotesize (color on line) (a)  Comparison of the transmission coefficient  $T(E)$ as a function of energy  in units of the potential height $V_0$ for the QB system  of the exact calculation using the transfer matrix method (solid  line) with resonance calculations
using Eq. (\ref{3}) for a number of poles: 10 (dashed line), 100 (dash-dot line) and 1000 (dotted line).  (b) A similar calculation for  the transmission amplitude ${\rm Re} {\bf t}(k) \textit{vs} {\rm Im} {\bf t}(k)$.}
\label{fAa}
\end{center}
\end{figure}

The set of complex poles of $\mathbf{t}(k)$ may be calculated using the Newton-Raphson method \cite{raphson}.
This method approximates a complex pole $\kappa_n$  by using  the iterative formula
\begin{equation}\label{2p}
\kappa_n^{r+1} \approx \kappa_n^{r} - \frac{t_{22}(\kappa_n^{r})}{t'_{22}(\kappa_n^{r})},
\end{equation}
where $ t'_{22}(k) = dt_{22}(k)/dk$. The approximate pole $\kappa_n^{r+1}$ goes into the exact pole,
at a given degree of accuracy, as the number of iterations increases.
In order to apply this method, it is necessary to provide an appropriate initial value for the approximate
pole $\kappa_n^0$.

In general for systems formed by a few alternating barriers and wells, as exemplified by Fig. \ref{f2}, the transmission coefficient \textit{vs} energy may be roughly characterized by three regimes: \textit{Regime I}, characterized by sharp isolated resonances (as in Fig. \ref{f2}(b)) or
groups of well defined overlapping resonances (as the resonance triplet in Fig. \ref{f2}(c)). This regime occurs usually for energies below the potential barrier height and refers to complex poles that are seated close to the real $k$-axis;  \textit{Regime II}, characterized by broad overlapping resonances. This regime is commonly found close to the potential barrier height and may extend up to energies $3$ or $4$ times the potential barrier height, as exemplified in all Figs. \ref{f2};
and \textit{Regime III}, involving much higher energies, well above the barrier height. There the transmission coefficient does not exhibit any appreciable resonance structure and just fluctuates very closely around unity.

There is in general no analytical expression for any initial approximate pole $\kappa_n^0$. An exception occurs along
the regime III, where there exists an asymptotic  formula for the location of complex poles which is valid for very large values of $n$  \cite{newton}
\begin{equation}
\kappa_n^0 \approx \frac{n\pi}{L} - i \frac{2}{L} \ln (n) + O(1); \,\,\,\,\, n \gg 1.
\label{2pa}
\end{equation}
One may substitute  Eq. (\ref{2pa}) into Eq. (\ref{2p}) to obtain the  pole $\kappa_n=\alpha_n-i\beta_n$ for
that very large value of $n$, say for example, $n=4000$.
Equation (\ref{2pa}) provides a relationship between the real parts of the $(n+1)$th and $(n-1)$th poles with the
$n$th pole
\begin{equation}\label{6p}
\alpha_{n\pm 1} \approx \alpha_n \pm \frac{\pi}{L} \equiv a_{n \pm 1},
\end{equation}
and for the corresponding imaginary parts,
\begin{equation}\label{6p2}
\beta_{n \pm 1} \approx \beta_n
\end{equation}
Hence one may write
\begin{equation}
\kappa^0_{n \pm 1} \approx \kappa_n \pm \Delta_r.
\label{2pb}
\end{equation}
where the step $\Delta_r$ is given by
\begin{equation}
\Delta_r = \frac{\pi}{L}.
\label{2pc}
\end{equation}
Then, one may calculate the  $(n-1)$th pole by substituting Eq. (\ref{2pb}) into the iterative Newton-Raphson formula
to evaluate the pole $\kappa_{n-1}$.  Repeating  this procedure successively allows to generate the poles for smaller values of $n$. Clearly this procedure permits also to obtain the poles for larger values of $n$. As the value of $n$ diminishes, however, on may reach  a situation
where, even if $n$ is still large, the iterative Newton-Raphson formula may fail.
We have found that in this circumstance Eq. (\ref{2pc}) still holds but Eq. (\ref{6p2}) becomes inaccurate.
In order to circumvent this situation one may proceed as follows. Once, as indicated above,  that it is determined that the pole $\kappa_n$ is asymptotic and has been calculated, one defines a rectangular region $I_{n-1}$ on the complex $k$ plane whose center contains the pole $\kappa^0_{n-1}$. This region is characterized by
\begin{eqnarray}
I_{n-1}&=&  [a_{n - 1} - \Delta_r/2, a_{n - 1} + \Delta_r/2] \times \nonumber \\[.3cm]
&&[-\beta_{n} - \Delta_i/2,-\beta_{n} + \Delta_i/2],
\label{region}
\end{eqnarray}
where $\Delta_i$  is a controllable parameter. Since the imaginary values of neighboring poles do not differ substantially,
it is sufficient to choose
\begin{equation}
\Delta_i= \beta_n.
\label{2pd}
\end{equation}
If, as indicated above,  the iterative formula given by Eq. (\ref{2p}) fails for a given initial value
$\kappa^0_{n-1}$, then a new initial value $\kappa^0_{n-1}$ is generated randomly according to the expression
\begin{equation}\label{7p}
\kappa_{n-1}^0 = \kappa_n - \Delta_r + \gamma_r\Delta_r + i \gamma_i\Delta_i,
\end{equation}
where, the parameters $\gamma_r$ and $\gamma_i$ are random numbers that vary, respectively, along the intervals
$-0.5\leq\gamma_r\leq0.5$ and $-0.5\leq\gamma_i\leq0.5$ to guarantee that the generated pole lies
within the region $I_{n-1}$. If the condition  $|t_{22}(\kappa_{n-1}^0)|<1$ is fulfilled, then the iterative formula (\ref{2p}) is applied. Otherwise or if the calculated pole lies outside $I_{n-1}$, that pole is disregarded and a new initial pole is generated
according to the above procedure. Usually, after a few random attempts convergence to a new pole is obtained.
If after many random attempts (M=$1000$ for the examples considered in this work) no convergence is achieved, that may suggest that
\textit{ Regime II} has been reached. This means that Eq. (\ref{2pc}) does not hold anymore. Then, it is convenient to define from that pole inwards thinner rectangular regions $I_{n-1}$. For the examples considered in this work, we choose $\Delta_r= \pi/20L$ and for $\Delta_i=2\beta_n$. Clearly, in this case some rectangular regions do not possess any poles. This procedure is capable to generate also the poles in \textit{Regime I}. Although in \textit{Regimes I} and \textit{II} the above procedure may generate repeated poles, a consequence that  Eq. (\ref{6p}) does not hold,
these poles may be easily identified and disregarded.
For \textit{Regime I} there is the alternative simple procedure to generate the initial values $\kappa_n^0$  by the rule of the half-width at half-maximum of the Breit-Wigner formula for the transmission coefficient.

Once a set of $N$ complex poles $\{\kappa_n\}$ has been  obtained, one may evaluate the transmission amplitude given by Eq. (\ref{3}), by  running it from $-N$ to $N$. One might then make a comparison of the resonance expansion, for different values of the number of poles, with the exact numerical calculation using the transfer matrix method \cite{ferry} to establish the appropriate number of poles for a given energy interval. Figure \ref{fAa}(a) provides a plot of the transmission coefficient \textit {vs} energy for the QB system discussed in the text for the exact numerical calculation using the transfer matrix method (solid line) and resonance expansions of ${\bf t}(k)$ for distinct number of poles: $N=10$ (dashed line), $N=100$ (dash-dot line) and $N=1000$ (dotted line). The energy interval extends up to $5$ times above the barrier height and one sees that as the number of poles increases the agreement with the exact calculation becomes better. Notice  that already with $N=100$ poles, the transmission coefficient is well reproduced for energies below the potential barrier height. Notice also that the calculation involving $1000$ poles is
still slightly different from the exact calculation in the interval  $ 4.0 < E/V_0 < 5.0$. The
calculation for the same system presented  in Fig. \ref{2}(c), that  involves $4000$ poles,
is indistinguishable from the exact calculation. One sees that away from sharp resonances, more resonance terms
are required to reproduce the exact calculation.
This is particularly striking in energy intervals where  $T(E)$  fluctuates very close to unity
where a very large number of resonance terms is necessary to reproduce the exact calculation.
Fortunately, very distant resonance poles are not difficult to calculate.
Figure \ref{fAa}(b) exhibits  similar calculations for the transmission amplitude. Here it is plotted ${\rm Re}\, {\bf t}(k) \textit{vs} \,{\rm Im}\, {\bf t}(k)$, to show that the resonance expansions of the transmission amplitude
become closer to the exact calculation as number of poles in the calculation increases.

\end{document}